\documentclass[lettersize,journal]{IEEEtran}
\usepackage{amsmath,amsfonts}
\usepackage{algorithmic}
\usepackage{algorithm}
\usepackage{array}
\usepackage[caption=false,font=normalsize,labelfont=sf,textfont=sf]{subfig}
\usepackage{textcomp}
\usepackage{stfloats}
\usepackage{url}
\usepackage{verbatim}
\usepackage{graphicx}
\usepackage{cite}
\usepackage{amssymb}
\usepackage{booktabs}
\usepackage{multirow}
\usepackage[hidelinks]{hyperref}
\begin{document}

\title{Unified Audio Generation and Editing via Joint Condition Modeling and Progressive Training}

\author{Haocheng~Dong, Yuheng~Lu, Cheng~Gong, Shansong~Liu, Xiao-Lei~Zhang, and~Xuelong~Li
\thanks{Work done during an internship at TeleAI.}
\thanks{Haocheng Dong is with the Department of Electronic Engineering and Information Science, University of Science and Technology
of China, China, and also with the Institute of Artificial Intelligence of China Telecom (TeleAI), China (e-mail: \mbox{haocheng-dong@mail.ustc.edu.cn}).}
\thanks{Yuheng Lu is with the Tianjin Key Laboratory of Cognitive Computing and Application, School of Artificial Intelligence, Tianjin University, China, and also with the Institute of Artificial Intelligence of China Telecom (TeleAI), China (e-mail: \mbox{luyuheng2024@tju.edu.cn}).}
\thanks{Cheng Gong, Shansong Liu, Xiao-Lei Zhang and Xuelong Li are with the Institute of Artificial Intelligence of China Telecom (TeleAI), China (e-mail: \mbox{\{gongc6, liuss31\}@chinatelecom.cn}, \mbox{xiaolei.zhang@nwpu.edu.cn}, \mbox{xuelong\_li@ieee.org}).}
}



\maketitle

\begin{abstract}
With the growing focus on audio in multimedia applications, numerous advanced works on audio generation have emerged. Existing studies typically treat text-to-audio (TTA) and other related audio generation tasks, such as instruction-based audio editing, as independent challenges, adopting task-specific architectures or modules. This absence of a unified modeling paradigm substantially increases the overhead and complexity of building a system for both audio generation and editing, while also leading to limited scalability. To address this issue, we introduce AudioWeave, a unified model for TTA and audio editing without additional task-specific components. Specifically, we propose a joint condition modeling approach with a factorized position embedding, enabling the diffusion transformer backbone to operate under heterogeneous inputs of TTA and audio editing. We further propose a progressive multistage training strategy to mitigate task competition and catastrophic forgetting caused by interference among multiple tasks. This in turn helps maintain the performance of each individual task and may even lead to improvements in certain aspects. Experimental results on TTA task and six audio editing tasks show that our unified model achieves competitive performance with task-specific models, laying a groundwork for further exploration of unified audio generation models.
\end{abstract}


\section{Introduction}
\IEEEPARstart{A}{udio} plays a crucial role in multimedia, serving not only as a modality for information delivery, but also as a medium for perceptual engagement. In diverse applications such as film, video games, and interactive media, audio significantly enriches user engagement and facilitates more impactful content communication, making audio generation a prominent topic of research in the field of AI-Generated Content (AIGC) \cite{foo2025ai, cao2025survey} and AI Flow \cite{an2026ai}.

With the demonstrated reliability of diffusion models \cite{ho2020denoising, song2021denoising} in generative tasks, a growing number of text-to-audio (TTA) generation models \cite{liu2023audioldm, huang2023make, haji2026taming, evans2025stable, hung2026tangoflux, li2025meanaudio} have emerged. These approaches have achieved remarkable performance on the TTA task, which aims to generate audio conditioned on text prompts. Generative audio editing has also become a highly focused direction within audio generation. Building on the capabilities of TTA generation models, several studies \cite{xu2024prompt, manor2024zero, jia2025audioeditor} perform audio editing via training-free approaches. Nevertheless, these methods lack support for instruction-based editing, thereby limiting their applicability. Some other approaches train dedicated instruction-based editing models \cite{wang2023audit, ungersbock2025sao, tao2025mmedit}, enabling flexible audio editing based on textual instructions. Since audio editing tasks require additional reference audio input, these corresponding methods are typically built with task-specific modules to handle it.

\begin{figure}[!t]
  \centering
  \includegraphics[width=\linewidth]{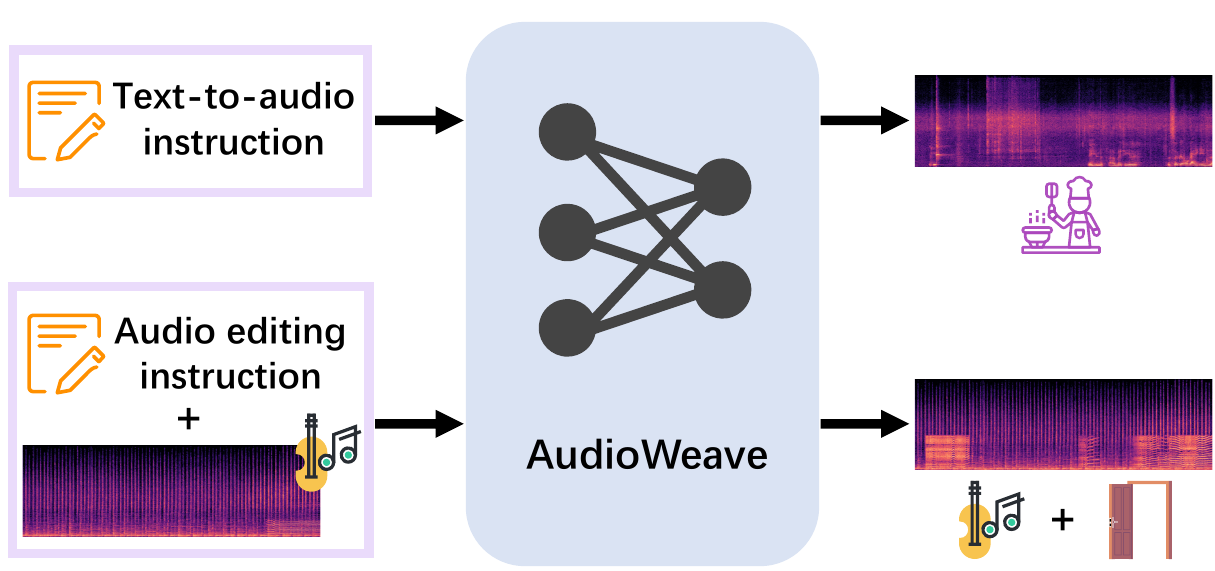}
  \caption{AudioWeave is a unified audio generation model capable of generating audio based on text prompts or editing audio following text instructions and reference audio.}
  \label{fig:intro}
\end{figure}

Due to the advancement of task-specific audio generation models, there has been an increasing interest in multifunctional AI tools capable of both audio generation and editing, in order to meet the growing demands of audio content creation. Audio-Agent \cite{wang2024audio} leverages a pretrained TTA model as a generation agent and employs a large language model (LLM) to decompose instructions, enabling audio generation and editing. However, relying only on instructions decomposed by an LLM to guide a single TTA model in handling multiple tasks can introduce inaccuracies and inconsistencies. Similarly, WavCraft \cite{liang2024wavcraft} utilizes an LLM to decompose user instructions, coordinating different models and audio processing functions for diverse audio generation and editing requirements. In these systems, each capability corresponds to a task-specific model or function, leading to increased system complexity and inefficiency. Moreover, they overlook the potential benefits that arise from interactions between tasks \cite{fu2025univg}, which could further improve overall performance.

The pursuit of general and flexible AI systems for multiple audio generation tasks has motivated researchers to explore unified models that can perform all tasks using a single architecture and a shared set of parameters. Such models aim to eliminate the need for task-specific architectures, reduce system complexity, and avoid developing elaborate workflows to support diverse functionalities. For example, UniAudio \cite{yang2024uniaudio} proposes an LLM-based model that unifies multiple audio generation tasks. However, the audio generation capability of autoregressive models based on discrete tokens does not match the performance of advanced latent diffusion models (LDMs) \cite{yang2025generative}. Works such as Audiobox \cite{vyas2023audiobox} and Fugatto \cite{valle2025fugatto} utilize flow matching \cite{lipman2023flow} models to unify diverse audio generation tasks, such as text-to-speech (TTS) and TTA, in a continuous representation space. Despite these advances, there is still a lack of sufficient research specifically dedicated to unified models for high-quality text-to-audio generation and faithful instruction-based audio editing.

To bridge this gap, as shown in Figure~\ref{fig:intro}, we propose AudioWeave, a unified model for generating audio conditioned on text prompts and editing audio based on text instructions and reference audio. To address the challenge of unifying tasks with heterogeneous conditioning inputs under a single model architecture and shared parameters, we propose a joint condition modeling approach with a factorized position embedding. This allows both audio generation and editing within a shared diffusion transformer (DiT) \cite{peebles2023scalable} backbone, enabling effective interactions among text, target audio, and optional reference audio.

Moreover, rather than adopting a joint training strategy, we propose a progressive multistage training strategy, which initially trains a base model on the TTA task, followed by a mix training on both TTA and audio editing with a task-specific attention mask. This training strategy facilitates achieving unified audio generation and editing capabilities while mitigating task competition and catastrophic forgetting caused by interference among multiple tasks.

The key contributions of our work are summarized as follows:
\begin{itemize}
    \item We propose a novel unified model for audio generation and editing, named AudioWeave.
    \item We propose a joint condition modeling approach with factorized position embedding, enabling a single diffusion transformer backbone to perform TTA and audio editing despite their different forms of input conditions.
    \item We introduce a progressive multistage training strategy with task-specific attention mask, enabling the model to effectively acquire TTA and audio editing capabilities, while also mitigating task competition and catastrophic forgetting.
    \item Experimental results suggest that our model achieves competitive performance compared to task-specific models on the TTA and six audio editing tasks, facilitating further exploration and applications of unified audio generation models.
\end{itemize}

The remainder of this paper is organized as follows. Section~\ref{sec:related} introduces related work on TTA and audio editing. Section~\ref{sec:method} presents the proposed AudioWeave in detail. Section~\ref{sec:setup} describes the experimental setup. Section~\ref{sec:result} reports and analyzes the experimental results. Finally, Section~\ref{sec:conclusion} concludes the paper.

\section{Related Work}
\label{sec:related}
\subsection{Text-to-Audio Generation}
The Text-to-Audio (TTA) generation task aims to synthesize audio from textual descriptions. With the rise of LDMs, many outstanding approaches have emerged in the field of TTA. AudioLDM \cite{liu2023audioldm} utilize a U-Net as the diffusion backbone and leverage contrastive language-audio pretraining (CLAP) model \cite{wu2023large, elizalde2023clap} as the pretrained encoder for conditioning. Similarly to AudioLDM, Make-An-Audio \cite{huang2023make} and TANGO \cite{majumder2024tango} both employ U-Net as their backbone. Unlike AudioLDM, which integrates conditional information via linear modulation layers \cite{perez2018film}, they interact with the text modality through a cross-attention mechanism.

AudioLCM \cite{10.1145/3664647.3681072} and ConsistencyTTA \cite{bai2024consistencytta} explore the application of consistency models \cite{song2023consistency} to audio generation, aiming to achieve competitive generative performance with fewer sampling steps. SoundCTM \cite{saito2024soundctm} is built upon the consistency trajectory model \cite{kim2024consistency} framework, where a teacher model is distilled to obtain a corresponding one-step or few-step audio generation model. IMPACT \cite{huang2025impact} proposes a mask-based audio generation model that operates in a continuous latent space and incorporates a diffusion head to enable iterative parallel decoding. GenAu \cite{haji2026taming} investigates the challenges of scaling data and models in audio generation. It proposes a scalable transformer architecture that is specially designed for the audio domain and additionally constructs a large-scale audio–text paired dataset. 

As Diffusion Transformers (DiT) \cite{peebles2023scalable} have demonstrated outstanding generative capabilities, works such as Stable Audio Open (SAO) \cite{evans2025stable} and EzAudio \cite{hai2025ezaudio} adapt DiT architectures in the TTA scenario. Instead of relying solely on adaptive layer norm blocks for text conditioning, they perform cross-attention with text embedding extracted by pretrained text encoders \cite{raffel2020exploring, chung2024scaling}. Inspired by FLUX \cite{flux2024}, TangoFlux \cite{hung2026tangoflux} adopts a similar architecture that combines multimodal diffusion transformer (MMDiT) \cite{esser2024scaling} blocks and DiT blocks to enhance multimodal interactions, while incorporating flow matching to improve both the efficiency and quality of generation. Recent research such as MeanAudio \cite{li2025meanaudio} has further optimized the generation efficiency on diffusion transformer architectures using methods such as MeanFlow \cite{geng2026mean}, further exploring the architectures and optimization strategies of TTA models. Drawing on these previous studies, we adopt a diffusion transformer trained with flow matching as the backbone of our model.

\begin{figure*}[!t]
  \centering  \includegraphics[width=\linewidth]{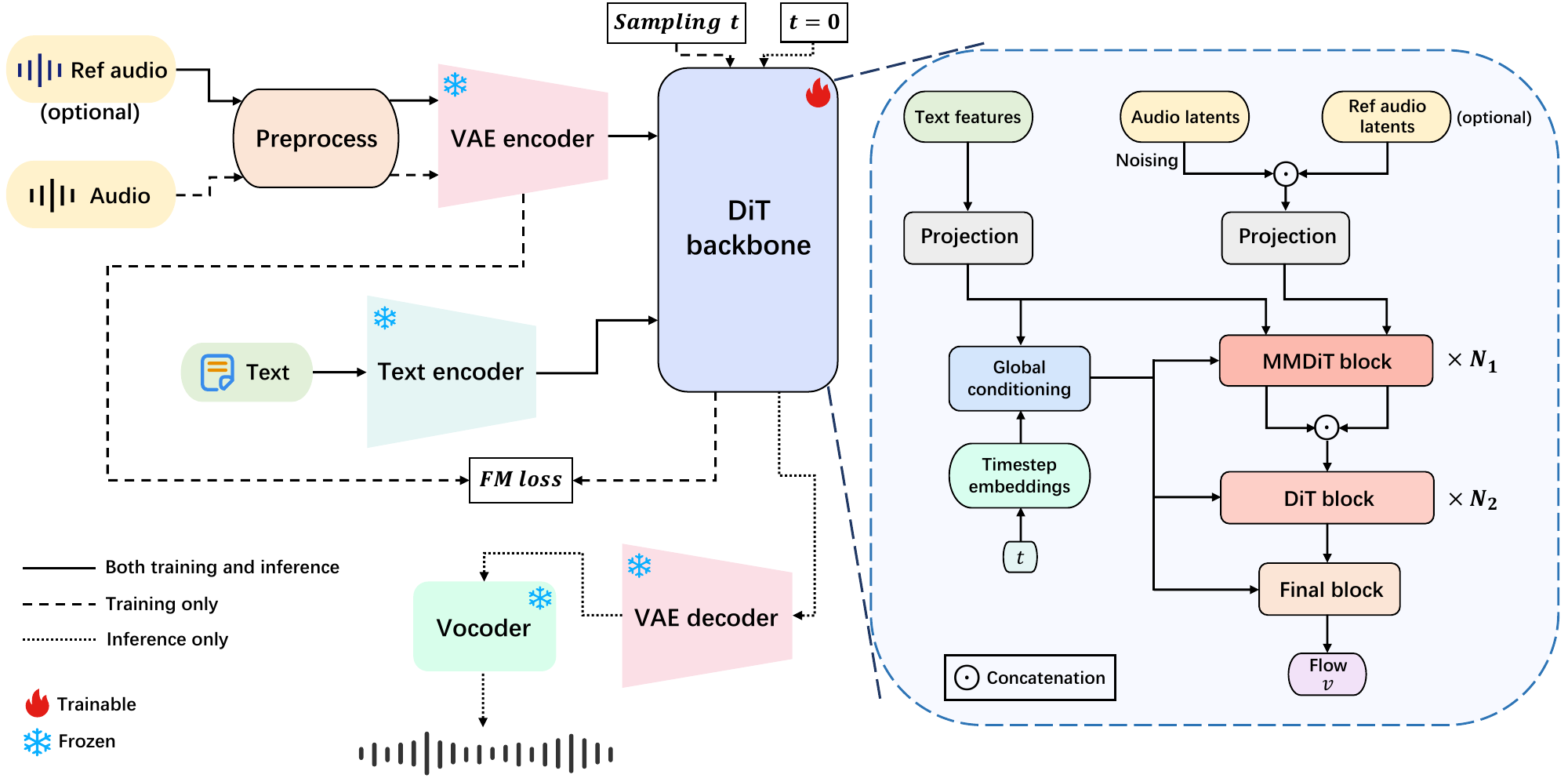}
  \caption{Overview of AudioWeave. The overall framework consists of an audio preprocessing module, an audio VAE, a text encoder, a vocoder, and a DiT backbone. The audio preprocessing module contains no trainable parameters. Only the DiT backbone is trainable, while all other components are pretrained. The proposed diffusion transformer backbone employs a hybrid architecture that combines dual-stream MMDiT blocks with single-stream DiT blocks.}
  \label{fig:overview}
\end{figure*}

\subsection{Audio Editing}
With the advancement of generative models for audio, researchers have increasingly turned their attention to their editing capabilities.
For training-free methods, ZETA and ZEUS \cite{manor2024zero} are built on edit-friendly DDPM inversion \cite{huberman2024edit}. ZETA implements text-based prompt-to-prompt editing while ZEUS provides unsupervised editing via perturbation control. AudioEditor \cite{jia2025audioeditor} employs DDIM inversion \cite{song2021denoising}, combined with Null-text Optimization \cite{mokady2023null} and end of tokens (EOT) suppression \cite{li2023get} to achieve reliable audio editing while maintaining similarity to the origin signal. AudioMorphix \cite{liang2025audiomorphix} integrates energy functions with attention manipulation, leveraging additional reference audio to enable fine-grained editing. However, these training-free methods primarily rely on the pretrained TTA models and full audio captions, which limits their ability to perform flexible instruction-based editing and consequently restricts their applicability.

Training-based audio editing models are capable of performing instruction-based audio editing. AUDIT \cite{wang2023audit} adopts a U-Net backbone with a cross-attention mechanism to enable instruction-based control. For the additional reference audio input, AUDIT extends the input channels for LDMs. This setup is commonly used in instruction-based audio editing models for reference audio conditioning. RFM-Editing \cite{gao2026rfm} leverages rectified flow matching to achieve more efficient and higher-quality editing performance. SAO-Instruct \cite{ungersbock2025sao} introduces a data construction pipeline for instruction-based audio editing, which combines three methodologies: Prompt-to-Prompt \cite{hertz2023prompttoprompt}, DDPM inversion \cite{manor2024zero}, and manual editing. SmartDJ \cite{lan2026smartdj} employs an audio language model as a planner to decompose user instructions to guide a stereo audio editing LDM. MMEdit \cite{tao2025mmedit} utilizes the cross-modal understanding capabilities of a large audio language model \cite{chu2024qwen2} to generate the semantic conditioning context. These training-based methods design dedicated modules specially tailored for audio editing tasks. Instead, our approach performs reference audio conditioning without modifying the existing model architecture.

\section{Proposed Method}
\label{sec:method}
\subsection{Overview}
We propose a unified model for audio generation and editing trained with a flow matching objective \cite{lipman2023flow, liu2023flow}. As shown in Figure~\ref{fig:overview}, the overall framework consists of an audio preprocessing module, a pretrained variational auto-encoder (VAE) \cite{cheng2025mmaudio}, a pretrained text encoder \cite{zhang2025encoder}, a pretrained vocoder \cite{lee2023bigvgan}, and a DiT backbone. With the exception of the DiT backbone, all other components are kept frozen during training.

For the DiT backbone, as presented in Figure~\ref{fig:overview}, we employ a hybrid design combining $N_1$ dual-stream MMDiT blocks and $N_2$ single-stream DiT blocks, and the
output flow is predicted by a final convolutional
block. Section~\ref{sec:dit} presents the details of the transformer blocks in our DiT backbone, and Section~\ref{sec:pe} presents the details of the proposed joint condition modeling with factorized position embedding.

During training, time $t$ is sampled from a logistic-normal distribution. At inference time, the DiT backbone is interpreted as an ODE solver that transports an initial noise sample at $t = 0$ to the final output at $t = 1$. We provide a detailed description of the proposed progressive multistage training strategy in Section~\ref{sec:train}.

\begin{figure*}[!t]
\centering
\subfloat[Dual-stream MMDiT block]{\includegraphics[width=0.50\linewidth]{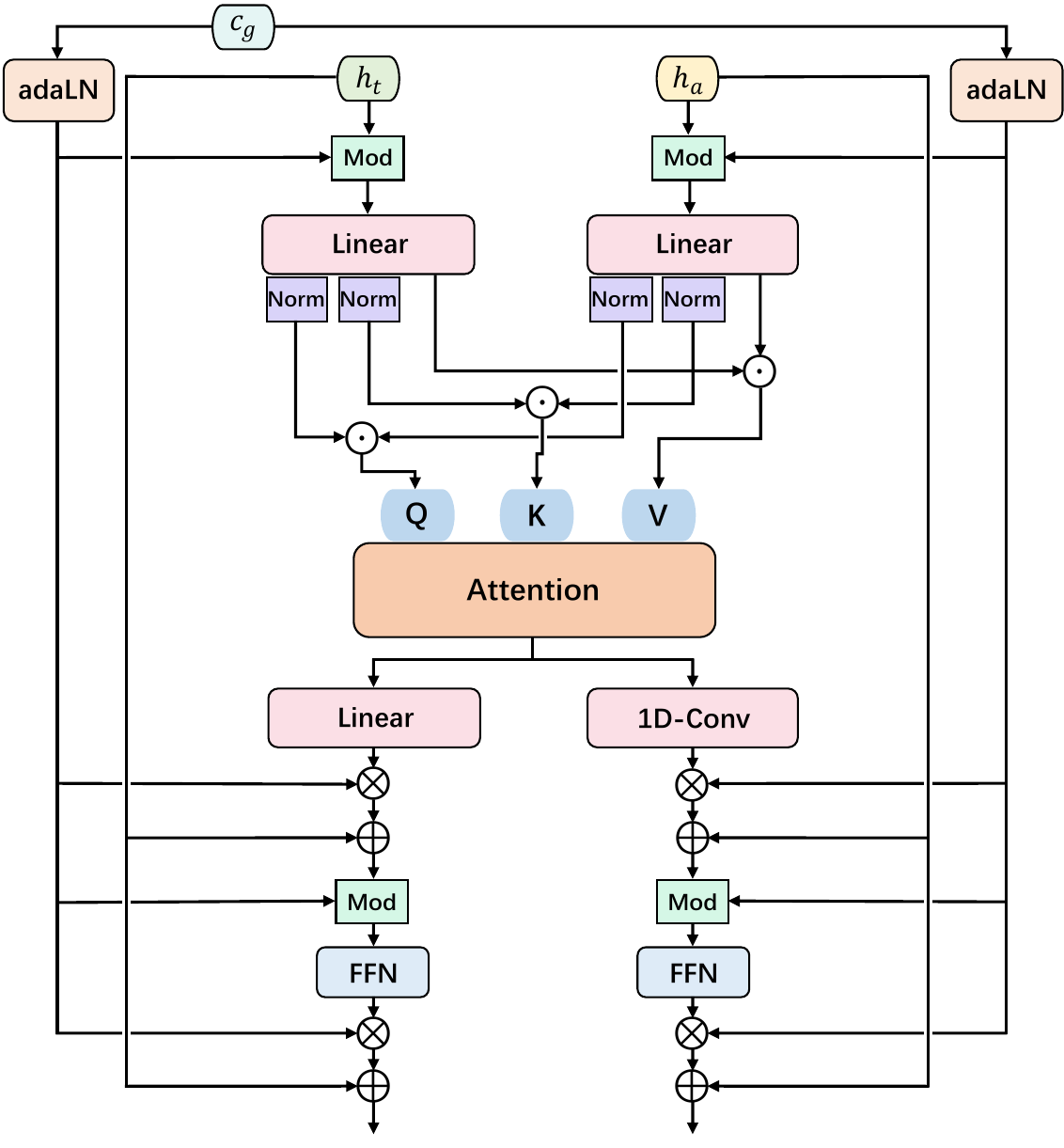}%
\label{fig:mb}}
\hfil
\subfloat[Single-stream DiT block]{\includegraphics[width=0.48\linewidth]{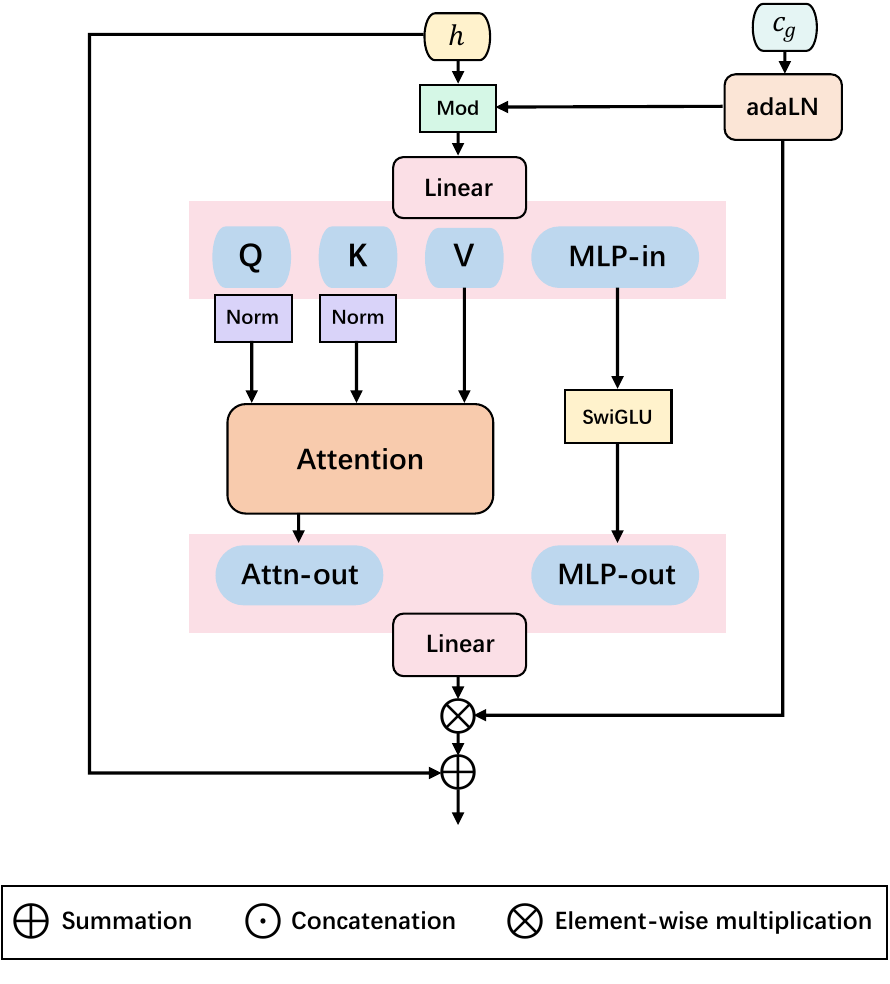}%
\label{fig:sb}}
\caption{Architectures of the dual-stream MMDiT block and the single-stream DiT block. (a) In the audio branch, the linear layers following joint attention and within the SwiGLU FFN are replaced with 1D convolutional layers, while the text branch retains the original configuration. $c_g$ denotes the global condition, $h_t$ denotes the hidden states of text stream, and $h_a$ denotes the hidden states of audio stream. (b) Parallel design of attention and FFN, rather than the standard architecture where attention is followed by a FFN. $c_g$ denotes the global condition, and $h$ denotes the fused hidden states.}
\label{fig:blocks}
\end{figure*}

\subsection{Diffusion Transformer Backbone}
\label{sec:dit}
\subsubsection{Dual-stream MMDiT block}
As illustrated in Figure~\ref{fig:mb}, we adapt the MMDiT block architecture from \cite{esser2024scaling} for audio generation, where the block consists of two separate branches corresponding to the text and audio streams. Given that 1D convolution is more capable of capturing the temporal structure \cite{cheng2025mmaudio}, in the audio stream branch, the linear layers following attention and within the SwiGLU feed-forward network (FFN) are replaced with 1D convolutional layers.

\subsubsection{Single-stream DiT block}
To enhance generation performance with deeper networks while maintaining parameter efficiency, existing work \cite{hung2026tangoflux, li2025meanaudio} employs a combination of DiT and MMDiT blocks rather than relying solely on MMDiT blocks. In line with this design for both effectiveness and efficiency, as shown in Figure~\ref{fig:sb}, we further improve our approach by adapting the DiT block from \cite{flux-2-2025} instead of the standard DiT block. Unlike the standard DiT block, this design requires only a single modulation of the hidden state, while the parallelization of attention and FFN enhances computational efficiency.

\begin{figure*}[!t]
  \centering  \includegraphics[width=\linewidth]{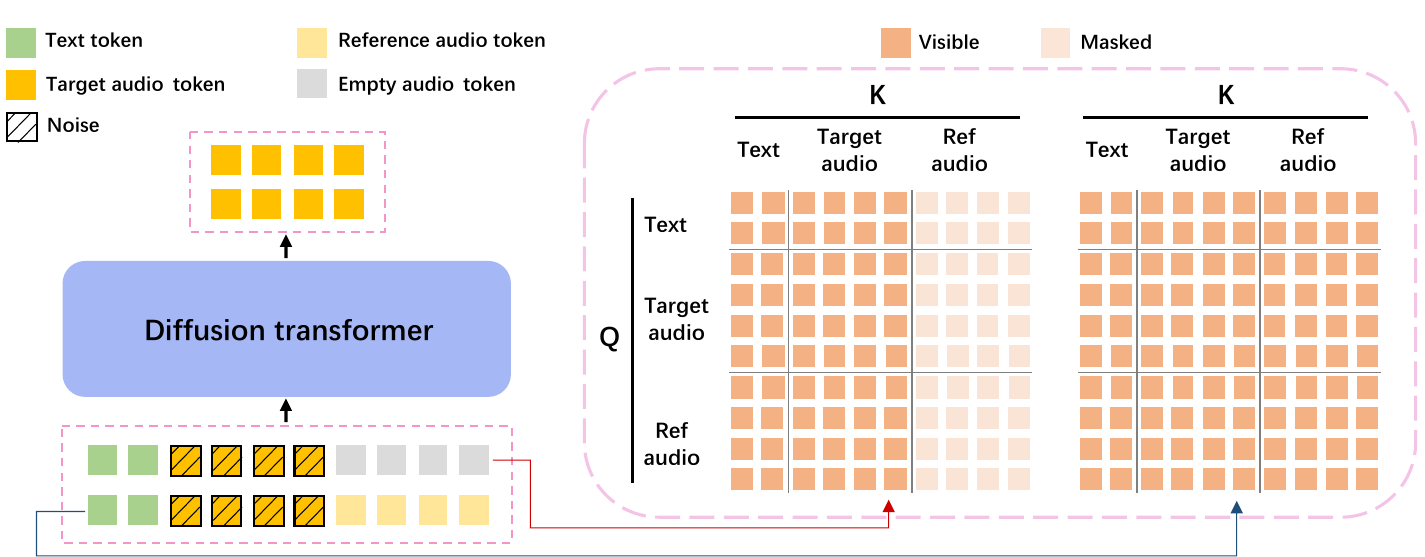}
  \caption{Mix training with task-specific attention mask. We mix the training data from TTA and audio editing tasks into a unified training set. For TTA training samples without reference audio, the corresponding reference sequence is replaced with empty audio tokens. The task-specific attention mask is designed to ensure that empty audio tokens are excluded from joint attention computation.}
  \label{fig:attnmask}
\end{figure*}

\subsection{Joint Condition Modeling}
\label{sec:pe}
A primary challenge of building a unified model for TTA and audio editing lies in the different forms of their input conditions. Expanding the input channels or introducing additional modules for the reference audio condition fails to yield a unified model.

To address this limitation, for audio editing tasks, we treat the reference audio condition as context and concatenate its latents $x_\text{ref}$ with the target audio latents $x_\text{tar}$ along the sequence dimension to form a joint representation $x_\text{audio}$:
\begin{equation}
    x_\text{audio} = \operatorname{Concat}(x_\text{tar}, x_\text{ref}),
\end{equation}
The resulting $x_\text{audio}$ is processed by the audio stream branch of the dual-stream MMDiT block. For the TTA task, we set $x_\text{audio} = x_\text{tar}$. Subsequently, in the single-stream DiT blocks, text and audio streams are sequence-wise concatenated into a fused stream for joint modeling:
\begin{equation}
    h_\text{fused} = \operatorname{Concat}(h_\text{text}, h_\text{audio}),
\end{equation}
where $h_\text{fused}$, $h_\text{text}$ and $h_\text{audio}$ denote the hidden states of the fused stream, the text stream, and the audio stream, respectively.

For both MMDiT and DiT blocks, the text sequence, optional reference audio sequence, and target audio sequence interact through a joint attention mechanism, making positional information crucial for capturing dependencies both across and within sequences. To this end, we propose a factorized position embedding for text, target audio, and reference audio tokens that encodes both inter-sequence and intra-sequence positional relationships. The proposed factorized position embedding can be formulated as:
\begin{equation}
    R^{d}(i,j)=
    \begin{bmatrix}
    R^{d_g}_{g}(i) & 0 \\
    0 & R^{d_l}_{l}(j)
    \end{bmatrix},
\end{equation}
where $i$ denotes the global position index of different sequences, $j$ denotes the local position index within a sequence, and the embedding dimension is decomposed as $d=d_g+d_l$. $R^{d_g}_{g}$ and $R^{d_l}_{l}$ are rotation matrices constructed following Rotary Position Embedding (RoPE) \cite{su2024roformer}, corresponding to the global inter-sequence and local intra-sequence positional components, respectively.

Based on the above formulation of factorized positional embedding, we introduce a position indexing scheme for text, target audio, and reference audio sequences. Since positional information among text tokens is already incorporated during text encoding and there is no explicit positional correspondence between each text token and the audio tokens, we assign a position of $(0, 0)$ to every token in the text sequence. For the $k$-th token in the target audio sequence, we assign a position of $(1, k)$. For the $n$-th token in the $m$-th reference audio sequence, we assign a position of $(m + 1, n)$. This factorized positional embedding distinguishes different sequences, preserves the semantic information of text tokens, and captures the temporal structure of audio tokens.

\subsection{Progressive Multistage Training}
\label{sec:train}
\subsubsection{Multistage Training Strategy}
TTA and audio editing can be formulated as distinct conditional generation problems, where the former is conditioned on text instructions, while the latter is conditioned on both editing instructions and reference audio. Moreover, audio editing tasks suffer from a scarcity of high-quality paired real-world datasets. Differences in input conditions and data distributions between tasks may induce task competition through direct joint training, making it difficult to achieve balanced generation and editing performance. To address this challenge, as presented in Algorithm~\ref{alg:train}, we propose a progressive multistage training strategy.

\begin{algorithm}[h]
\caption{Progressive Multistage Training.}\label{alg:train}
\begin{algorithmic}
\STATE
\STATE \textbf{Input:} datasets $D_\text{TTA}$ and $D_\text{edit}$, steps $S_1$ and $S_2$
\STATE \textbf{Initialize} model parameters $\theta$
\FOR{$step=0$ to $S_1$}
\STATE \textbf{randomly select} batch $B \subset D_\text{TTA}$
\STATE $(x_1, c_\text{text}) \leftarrow B$
\STATE \textbf{sample} $t$ and $x_0$
\STATE $x_t \leftarrow tx_1 + (1-t)x_0$
\STATE $v_t \leftarrow x_1 - x_0$
\STATE $    {\mathcal{L_{\text{FM}}}} \leftarrow \|v_{\theta}(t, x_t, c_\text{text}) - v_t \|_2^2$
\STATE \textbf{update} $\theta$ via $\nabla_{\theta} \mathcal{L_{\text{FM}}}$
\ENDFOR
\FOR{$step=0$ to $S_2$}
\STATE \textbf{randomly select} batch $B \subset D_\text{TTA} \cup D_\text{edit}$
\STATE $(x_1, c_\text{text}, c_\text{ref}) \leftarrow B$
\STATE \textbf{sample} $t$ and $x_0$
\STATE $x_t \leftarrow tx_1 + (1-t)x_0$
\STATE $v_t \leftarrow x_1 - x_0$
\STATE $    {\mathcal{L_{\text{FM}}}} \leftarrow \|v_{\theta}(t, x_t, c_\text{text}, c_\text{ref}) - v_t \|_2^2$
\STATE \textbf{update} $\theta$ via $\nabla_{\theta} \mathcal{L_{\text{FM}}}$
\ENDFOR
\STATE \textbf{return} $\theta$
\end{algorithmic}
\end{algorithm}

\textbf{Stage 1: Base model training.} At this stage, we train the model from scratch dedicated to TTA, serving as the foundation for subsequent training.

\textbf{Stage 2: Mix training.} Building on the base model in the first training stage, we conduct mix training in this stage. To mitigate catastrophic forgetting, we mix the training datasets of TTA and audio editing. For TTA training samples without reference audio, a fixed value is used to represent the empty audio token. This empty audio token is applied to construct an empty reference audio sequence to ensure that all training samples share the same format and sequence length.

\subsubsection{Task-specific Attention Mask}
\label{subsubsec:attnmask}
We employ a task-specific attention mask based on task identity during the mix training stage. For TTA training samples, as shown in Figure~\ref{fig:attnmask}, all query tokens are prevented from attending to key tokens of the empty reference audio, while all other tokens remain bidirectionally accessible. For audio editing training samples, full bidirectional attention is enabled for all tokens. This ensures consistency of the condition format between training and inference for the TTA and editing tasks.

\subsubsection{Classifier-free Guidance}
We adopt a stochastic condition dropout strategy during all training stages to enable classifier-free guidance (CFG) \cite{ho2021classifierfree} during inference. Specifically, we randomly set the text or reference audio to empty conditions with predefined probabilities. For the textual modality, the empty condition is represented by a pre-extracted empty feature, which remains fixed throughout all training stages. In contrast, the empty condition for the reference audio is modeled as a learnable token that is optimized during mix training. Since different tasks require different conditioning inputs, we employ task-specific formulations of CFG. During inference, for the TTA task, we have:
\begin{equation}
\label{eq:ttacfg}
    v_{\theta} = v_{\theta}(t, x_t, \varnothing_\text{text}) + \omega_0(v_{\theta}(t, x_t, c_\text{text})-v_{\theta}(t, x_t, \varnothing_\text{text})),
\end{equation}
for audio editing tasks, we have:
\begin{equation}
\label{eq:mtcfg}
\begin{split}
    v_{\theta} = & \ v_{\theta}(t, x_t, \varnothing_\text{text}, \varnothing_\text{ref})\\ & + \omega_1(v_{\theta}(t, x_t, \varnothing_\text{text}, c_\text{ref})-v_{\theta}(t, x_t, \varnothing_\text{text}, \varnothing_\text{ref}))\\ & + \omega_2(v_{\theta}(t, x_t, c_\text{text}, c_\text{ref})-v_{\theta}(t, x_t, \varnothing_\text{text}, c_\text{ref})),
\end{split}
\end{equation}
where $c_\text{text}$ denotes the text condition; $c_\text{ref}$ denotes the reference audio condition; $\varnothing_\text{text}$ and $\varnothing_\text{ref}$ are the empty conditions of text and reference audio, respectively; $t$ is the timestep; $x_t$ represents the input latents at time $t$; and $\omega_0$ and $(\omega_1, \omega_2)$ are the guidance scales.

\section{Experimental Setup}
\label{sec:setup}
\subsection{Datasets}
\subsubsection{Text-to-audio datasets}
For the base model training, we utilize multiple audio datasets, including AudioCaps \cite{audiocaps}, AudioSet \cite{gemmeke2017audio}, WavCaps \cite{mei2024wavcaps} and YouTube-8M \cite{abu2016youtube}. Each training sample consists of a 10-second audio clip paired with its corresponding textual caption. For audio segments shorter than 10 seconds, we discard them. For audio recordings longer than 10 seconds, we extract up to two non-overlapping 10-second segments.

\textbf{AudioSet.} AudioSet is a large-scale, manually annotated dataset of audio events, consisting of an ontology of 632
audio classes and over 2 million audio clips collected from YouTube videos. We selected a subset of approximately 254K audio clips that maintains the balance of the audio classes as much as possible for training.

\textbf{YouTube-8M.} YouTube-8M is a large-scale multi-label video classification dataset comprising approximately 8 million videos. We selected a subset comprising approximately 279K audio clips, likewise striving to maintain class balance as much as possible.

For both subsets of AudioSet and YouTube-8M, we use Gemini~2.5 \cite{comanici2025gemini} to generate captions for audio instead of relying on the corresponding labels.

\textbf{AudioCaps.} AudioCaps is a subset of AudioSet and contains about 50K audio clips paired with human-written captions. We employ the official training split for model training.

\textbf{WavCaps.} The WavCaps dataset contains approximately 400K audio clips of varying durations, each paired with captions generated by LLMs. Since some audio clips in WavCaps can be segmented into two 10-second segments, we assign the original caption to one segment, while for the other segment, we generate a caption using Gemini~2.5. We ultimately obtained about 400K audio-text pairs.

We evaluate the performance of TTA generation using the AudioCaps test split which contains 975 audio clips, each paired with five distinct captions. As in previous work \cite{li2025meanaudio, liu2023audioldm}, we randomly select one caption for each audio clip as the text instruction during evaluation. Information on training and test sets for the TTA task is summarized in Table~\ref{tab:ttadata}.

\begin{table}[!t]
    \caption{Summary of TTA datasets, including data sources and sizes of training and test sets.}
    \label{tab:ttadata}
    \centering
    \begin{tabular}{l|ccccc}
        \toprule
        \textbf{Split} & AudioCaps & AudioSet & WavCaps & YouTube-8M & Total \\
        \midrule
        Train & 43K & 254K & 400K & 279K & 976K \\
        Test & 975 & $-$ & $-$ & $-$ & 975 \\
        \bottomrule
  \end{tabular}
\end{table}

\begin{table}[!t]
    \caption{Details of audio editing datasets, including example instructions, and data scale of the training and test sets for six editing tasks.}
  \label{tab:editdataset}
    \centering
  \resizebox{\columnwidth}{!}{
  \begin{tabular}{l|l|cc}
    \toprule
    \textbf{Task} & \textbf{Example Instruction} & \textbf{Training Set} & \textbf{Test Set} \\
    \midrule
    Adding & Add \{\} at the beginning. & 160K & 1K \\
    Removing & Remove \{\}. & 200K & 1K \\
    Replacement & Replace \{\} with \{\}. & 160K & 1K \\
    Reordering & Reorder \{\} and \{\}. & 120K & 1K \\
    Inpainting & Inpaint the missing part. & 120K & 1K \\
    SR & Perform audio super-resolution. & 120K & 1K \\
    \bottomrule
  \end{tabular}}
\end{table}

\begin{figure}[!t]
  \centering  \includegraphics[width=\linewidth]{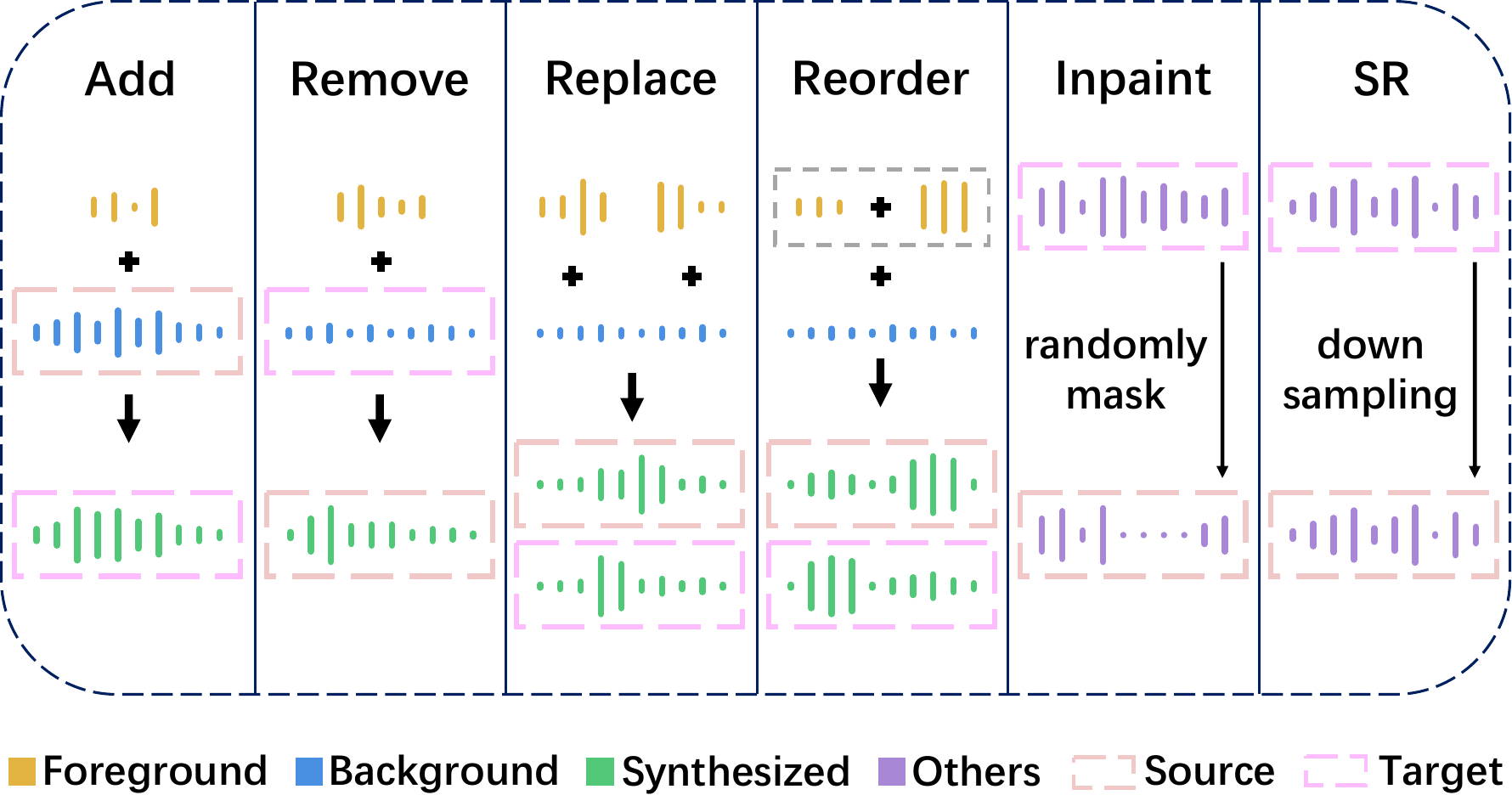}
  \caption{The construction of audio editing datasets. \textit{Source} means the input reference audio of editing, and \textit{Target} means the target output of editing. \textit{Others} denote the setting where each target–source audio pair is constructed from a single audio clip.}
  \label{fig:editdata}
\end{figure}

\subsubsection{Audio editing datasets}
Following prior works \cite{wang2023audit, ungersbock2025sao, tao2025mmedit}, we apply a similar approach of manually synthesizing edited data and use comparable instruction formats to construct audio editing task datasets on AudioSet, ESC-50 \cite{piczak2015esc}, Nonspeech7k \cite{rashid2023nonspeech7k}, UrbanSound8K \cite{salamon2014dataset} and FSD50K \cite{fonseca2021fsd50k}.

\textbf{ESC-50.} The ESC-50 dataset consists of 2K labeled environmental audio recordings, evenly distributed across 50 distinct classes.

\textbf{Nonspeech7k. } The Nonspeech7k dataset is a collection of human non-speech sounds comprising seven distinct categories, with a total of 7,014 audio clips.

\textbf{UrbanSound8K. } UrbanSound8K comprises ten annotated classes of urban sounds collected from Freesound \cite{font2013freesound}, which contains 8,732 labeled audio segments.

\textbf{FSD50K.} FSD50K contains more than 51K audio clips gathered from Freesound and has been manually annotated across 200 categories derived from the AudioSet ontology.

We select a total of six distinct audio editing tasks, including: adding, removing, replacement, reordering, inpainting and super-resolution (SR). As presented in Figure~\ref{fig:editdata}, for the tasks of adding, removing, replacement, and reordering, a random audio clip is sampled from a randomly selected audio category as the foreground. Specifically, the foreground audio is drawn exclusively from ESC-50, Nonspeech7k, UrbanSound8K and FSD50K datasets. Meanwhile, the background audio is randomly chosen from an AudioSet balanced subset, which contains about 43K audio clips. For the inpainting and SR tasks, audio are randomly sampled from the Audioset subset utilized in TTA task. Since there is no standardized benchmark for audio editing, we randomly sample 10\% of each raw dataset to build test sets, while the remaining is used to build editing training sets. More detailed information is provided in Table~\ref{tab:editdataset}.

\begin{table}[!t]
    \caption{Details of the baselines. The parameter count includes only the backbone parameters. In particular, we report number of function evaluations (NFE) as the number of denoising steps.}
  \label{tab:baseline}
    \centering
    \resizebox{\columnwidth}{!}{
  \begin{tabular}{l|ccccccc}
    \toprule
    \textbf{Model} & \#Params & Backbone & Conditioning Encoder & NFE \\
    \midrule
    \textit{TTA Models} \\
    AudioLDM 2 & 750M & U-Net & CLAP + FLAN-T5 \cite{chung2024scaling} & 200 \\
    Tango 2 & 866M & U-Net & FLAN-T5 & 200 \\
    Make-An-Audio 2 & 160M & Transformer & CLAP + T5 \cite{raffel2020exploring} & 100 \\
    Stable Audio Open & 1057M & Transformer & T5 & 100 \\
    EzAudio & 874M & Transformer & FLAN-T5 & 100 \\
    GenAu & 1.25B & Transformer & CLAP + FLAN-T5 & 200 \\
    TangoFlux & 515M & Transformer & FLAN-T5 & 50 \\
    MeanAudio & 480M & Transformer & CLAP + FLAN-T5 & 25 \\
    \midrule
    \textit{Audio Editing Models} \\
    SAO-Instruct & 1057M & Transformer & T5 & 100 \\
    MMEdit & 616M & Transformer & Qwen2-Audio \cite{chu2024qwen2} & 50 \\
    \bottomrule
  \end{tabular}}
\end{table}

\subsection{Metrics}
\subsubsection{Objective Metrics}
For text-to-audio generation, the evaluation is conducted using standard objective metrics: Fréchet Distance (FD), Kullback-Leibler (KL) divergence, Inception Score (IS) \cite{salimans2016improved}, and CLAP score. FD evaluates the fidelity and diversity by measuring the distance between feature distributions of generated and real audio. KL is used to quantify the difference between the probability distributions of synthesized and real audio. IS reflects both audio quality and diversity using a pretrained classifier. CLAP score measures semantic alignment between audio and text by computing the cosine similarity between text and audio embeddings.

For FD, audio embeddings are extracted using PANNs \cite{kong2020panns} ($\text{FD}_{\text{PANNs}}$) and PaSST \cite{koutini2022efficient} ($\text{FD}_{\text{PaSST}}$), which are also adopted as classifiers for KL divergence ($\text{KL}_{\text{PANNs}}$ and $\text{KL}_{\text{PaSST}}$) and IS ($\text{IS}_{\text{PaSST}}$ and $\text{IS}_{\text{PaSST}}$). For CLAP score, we employ a base CLAP model\footnote{\url{https://huggingface.co/lukewys/laion_clap/blob/main/630k-audioset-fusion-best.pt}} denoted by $\text{CLAP}_\text{base}$ and a larger-capacity CLAP model\footnote{\url{https://huggingface.co/lukewys/laion_clap/blob/main/music_speech_audioset_epoch_15_esc_89.98.pt}} trained on a larger-scale dataset denoted by $\text{CLAP}_\text{large}$. In the following, the subscripts associated with each metric explicitly denote the model used for its computation.

For audio editing, following previous studies \cite{wang2023audit, ungersbock2025sao}, we chose FD and KL divergence to evaluate the distribution matching, and use log spectral distance (LSD) to quantify the distance between frequency spectrograms.

\subsubsection{Subjective Metrics}
For all tasks, we adopt the mean opinion score (MOS) on a discrete scale ranging from 1 to 5 for human evaluation. The assessment is conducted on two dimensions: quality (MOS-Q) and relevance (MOS-R).

For text-to-audio, MOS-Q evaluates the general audio quality, such as whether the generated audio sounds natural, without considering semantic alignment with text. In contrast, MOS-R measures the semantic alignment between the generated audio and the textual prompt. We randomly select 100 audio samples from the AudioCaps test set and have them rated by 18 researchers in audio signal processing.

For audio editing, similar to TTA, MOS-Q is used to evaluate the perceptual quality of audio, including whether the unedited parts are faithfully preserved and the generation quality of the edited regions. MOS-R measures the degree to which the editing adheres to the given textual instructions. We randomly select 15 audio samples from the test set for each task, and subjective evaluations were conducted by the same group of 18 researchers.

\subsection{Baselines}
We select state-of-the-art (SOTA) task-specific models as baselines to compare performance with our model on TTA and editing tasks.
\subsubsection{Text-to-audio}
We consider eight representative and recent SOTA TTA models as our baselines, including AudioLDM~2 \cite{audioldm2-2024taslp}, Tango~2 \cite{majumder2024tango}, Make-An-Audio~2 \cite{huang2023makeanaudio}, Stable Audio Open \cite{evans2025stable}, EzAudio \cite{hai2025ezaudio}, GenAu \cite{haji2026taming}, TangoFlux \cite{hung2026tangoflux} and MeanAudio \cite{li2025meanaudio}. Detailed information for each baseline is provided in Table~\ref{tab:baseline}. For all baselines, inference is performed using the official implementations and checkpoints from their publicly released code repositories.

\subsubsection{Audio editing} Since our model focuses on instruction-based audio editing, whereas training-free methods require paired original and target captions, we do not consider them as baselines. We select two recent SOTA instruction-based audio editing models, SAO-Instruct \cite{ungersbock2025sao} and MMEdit \cite{tao2025mmedit} as baselines. We adopt the publicly available official implementations and checkpoints for both baselines; further details are provided in Table~\ref{tab:baseline}. Following the commonly adopted setting from previous works \cite{wang2023audit, gao2026rfm, ungersbock2025sao, tao2025mmedit, lan2026smartdj}, where the input channel is extended for reference audio input, we additionally train a baseline named DiT-Edit, utilizing an input projection module specialized for audio editing tasks.

\begin{table*}[!t]
    \caption{Objective and subjective evaluation results for text-to-audio generation on AudioCaps test set. The best results are marked in bold, and the second-best results are underlined. Due to the high cost of human evaluation, subjective scoring is performed on three baselines: GenAu, TangoFlux, and MeanAudio.}
  \label{tab:TTA}
    \centering
    \resizebox{\textwidth}{!}{
  \begin{tabular}{l|cc|cc|cc|cc|cc}
    \toprule
    \textbf{Model} & $\text{FD}_{\text{PANNs}}$ $\downarrow$ & $\text{FD}_{\text{PaSST}}$ $\downarrow$ & $\text{KL}_{\text{PANNs}}$ $\downarrow$ & $\text{KL}_{\text{PaSST}}$ $\downarrow$ & $\text{IS}_{\text{PANNs}}$ $\uparrow$ & $\text{IS}_{\text{PaSST}}$ $\uparrow$ & $\text{CLAP}_\text{base}$ $\uparrow$ & $\text{CLAP}_\text{large}$ $\uparrow$ & \text{MOS-Q} $\uparrow$ & \text{MOS-R} $\uparrow$\\
    \midrule
    AudioLDM 2 & 44.744 & 423.714 & 1.748 & 1.647 & 6.720 & 5.606 & 0.369 & 0.201 & $-$ & $-$ \\
    Tango 2 & 14.816 & 249.518 & \textbf{1.138} & \textbf{0.972} & 10.534 & 7.266 & \underline{0.536} & 0.330 & $-$ & $-$ \\
    Make-An-Audio 2 & 15.245 & 183.782 & 1.228 & 1.293 & 9.753 & 10.209 & 0.476 & 0.295 & $-$ & $-$ \\
    Stable Audio Open & 41.974 & 273.215 & 2.293 & 2.067 & 9.864 & 8.933 & 0.281 & 0.212 & $-$ & $-$ \\
    EzAudio & 14.431 & \textbf{126.971} & 1.216 & 1.149 & 11.619 & 10.050 & 0.519 & 0.376 & $-$ & $-$ \\
    GenAu & 14.455 & \underline{127.201} & 1.296 & 1.145 & \textbf{13.920} & \underline{11.380} & 0.526 & 0.329 & 3.602 & 3.575 \\
    TangoFlux & 19.468 & 202.190 & 1.156 & \underline{0.987} & \underline{13.075} & 9.120 & \textbf{0.544} & 0.317 & 3.466 & 3.868 \\
    MeanAudio & \textbf{12.587} & 152.210 & \underline{1.153} & 1.204 & 12.302 & \textbf{12.425} & 0.534 & 0.339 & 3.642 & 3.736 \\
    \midrule
    Ours-TTA-only & 14.431 & 134.354 & 1.199 & 1.096 & 12.993 & 10.648 & 0.528 & \textbf{0.427} & \underline{3.752} & \underline{3.891} \\
    Ours-Full & \underline{14.288} & 134.341 & 1.282 & 1.171 & 12.168 & 9.734 & 0.513 & \underline{0.418} & \textbf{3.805} & \textbf{3.977} \\
    \bottomrule
  \end{tabular}}
\end{table*}

\begin{table*}[!t]
    \caption{Objective and subjective evaluation results for six editing tasks. The best results are marked in bold.}
  \label{tab:edit}
    \centering
  \begin{tabular}{l|l|cc|cc|c|cc}
    \toprule
    \textbf{Task} & \textbf{Model} &  $\text{FD}_{\text{PANNs}}$ $\downarrow$ & $\text{FD}_{\text{PaSST}}$ $\downarrow$ & $\text{KL}_{\text{PANNs}}$ $\downarrow$ & $\text{KL}_{\text{PaSST}}$ $\downarrow$ & $\text{LSD}$ $\downarrow$ & \text{MOS-Q} $\uparrow$ & \text{MOS-R} $\uparrow$\\
    \midrule
    \multirow{5}{*}{Adding}
    & SAO-Instruct & 15.831 & 229.516 & 1.234 & 1.419 & 1.993 & 3.361 & 1.667 \\
    & MMEdit & 17.005 & 217.337 & 1.477 & 1.413 & 6.383 & 2.917 & 3.028 \\
    & DiT-Edit & 10.033 & 99.942 & 0.632 & 0.566 & 1.386 & 3.889 & 4.222 \\
    & Ours-Edit-only & 9.950 & \textbf{97.456} & \textbf{0.596} & \textbf{0.537} & 1.349 & \textbf{4.028} & \textbf{4.389} \\
    & Ours-Full & \textbf{9.602} & 98.824 & 0.609 & 0.560 & \textbf{1.343} & 3.778 & 4.333 \\
    \midrule
    \multirow{5}{*}{Removing}
    & SAO-Instruct & 28.583 & 245.519 & 1.262 & 1.082 & 1.842 & 3.222 & 1.685 \\
    & MMEdit & 34.274 & 284.490 & 2.144 & 1.827 & 6.826 & 3.000 & 2.704 \\
    & DiT-Edit & \textbf{10.783} & \textbf{93.880} & \textbf{0.324} & \textbf{0.216} & \textbf{1.463} & \textbf{4.352} & \textbf{4.481} \\
    & Ours-Edit-only & 10.834 & 94.722 & \textbf{0.324} & 0.217 & 1.464 & 4.204 & 4.407 \\
    & Ours-Full & 11.335 & 97.392 & 0.337 & 0.222 & 1.465 & 4.222 & 4.444 \\
    \midrule
    \multirow{5}{*}{Replacement}
    & SAO-Instruct & 22.497 & 203.293 & 1.650 & 1.446 & 1.843 & 2.806 & 1.694 \\
    & MMEdit & 34.626 & 300.542 & 3.813 & 2.720 & 6.373 & 2.472 & 2.500 \\
    & DiT-Edit & \textbf{10.280} & \textbf{96.594} & \textbf{0.524} & \textbf{0.499} & 1.314 & 3.889 & 4.250 \\
    & Ours-Edit-only & 10.548 & 103.357 & 0.554 & 0.519 & 1.342 & 3.917 & 4.250 \\
    & Ours-Full & 10.470 & 103.291 & 0.551 & 0.524 & \textbf{1.287} & \textbf{4.139} & \textbf{4.306} \\
    \midrule
    \multirow{5}{*}{Reordering}
    & SAO-Instruct & 48.623 & 409.605 & 3.319 & 2.467 & 2.585 & 2.111 & 1.944 \\
    & MMEdit & 23.216 & 275.217 & 1.487 & 1.222 & 6.506 & 2.944 & 2.750 \\
    & DiT-Edit & 7.820 & 73.193 & 0.350 & 0.243 & 1.063 & 4.194 & \textbf{4.417} \\
    & Ours-Edit-only & \textbf{7.477} & \textbf{71.727} & \textbf{0.324} & \textbf{0.229} & \textbf{1.046} & \textbf{4.222} & \textbf{4.417} \\
    & Ours-Full & 8.318 & 79.709 & 0.349 & 0.258 & 1.053 & 4.194 & 4.389 \\
    \midrule
    \multirow{4}{*}{Inpainting}
    & SAO-Instruct & 27.413 & 221.924 & 0.776 & 0.661 & 3.290 & 2.611 & 2.028 \\
    & DiT-Edit & 14.012 & 112.337 & 0.486 & 0.356 & \textbf{1.300} & 3.722 & 3.722 \\
    & Ours-Edit-only & 12.861 & 97.719 & 0.460 & 0.309 & 1.326 & 3.750 & 3.694 \\
    & Ours-Full & \textbf{11.695} & \textbf{89.320} & \textbf{0.430} & \textbf{0.275} & 1.330 & \textbf{3.889} & \textbf{3.861} \\
    \midrule
    \multirow{4}{*}{Super-resolution}
    & SAO-Instruct & 13.264 & 140.932 & 0.440 & 0.413 & 2.449 & 3.250 & 3.444 \\
    & DiT-Edit & 7.831 & 69.697 & 0.245 & 0.183 & \textbf{1.426} & \textbf{3.861} & \textbf{4.000} \\
    & Ours-Edit-only & \textbf{7.825} & \textbf{69.027} & \textbf{0.244} & \textbf{0.176} & 1.435 & \textbf{3.861} & 3.944 \\
    & Ours-Full & 8.675 & 78.782 & 0.279 & 0.210 & 1.540 & \textbf{3.861} & 3.917 \\
    \bottomrule
  \end{tabular}
\end{table*}

\subsection{Implementation Details}
We set the audio sampling rate to 44.1 kHz. We adopt the pretrained audio VAE for 44.1kHz audio from MMAudio \cite{cheng2025mmaudio}. Its latent representation has 40 channels and the overall temporal downsampling rate is 1024. For the vocoder, we adopt BigVGAN-v2 \cite{lee2023bigvgan}, which supports a sampling rate of 44.1 kHz. For text conditioning, we use the text encoder of T5Gemma \cite{zhang2025encoder}, with an embedding dimension of 1024 and a standardized sequence length of 77 tokens. The DiT model consists of 8 MMDiT blocks and 16 DiT blocks. The number of attention heads is set to 14, and the total dimension of the hidden states is 896. The entire DiT backbone comprises approximately 672M parameters.

For DiT model training, we utilize 4 NVIDIA H100 GPUs. For the multistage training setup, the base model training stage was trained for 500K steps, while the mix training stage was trained for 400K steps, both with a batch size of 128. Throughout all training stages, we use the AdamW optimizer with a learning rate of $1 \times 10^{-4}$, along with a linear warm-up schedule in the first 1K steps. During inference, we set the sampling steps to 50 and set the CFG scale as $\omega_0=4.5$, $\omega_1=1.5$, and $\omega_2=4.0$.

\section{Results and Analysis}
\label{sec:result}
In this section, we conduct a thorough and comprehensive evaluation of our model on the TTA task, as well as six audio editing tasks. Several samples are shown on the demo webpage\footnote{\url{https://haochengdong.github.io/AudioWeave_Demo/}}.

\begin{table*}[!t]
    \caption{Ablation study on training strategies and position embedding. The best results are marked in bold.}
  \label{tab:ablation}
    \centering
  \resizebox{\textwidth}{!}{
  \begin{tabular}{l|l|cc|cc|cc|cc|c}
    \toprule
    \textbf{Task} & \textbf{Model} & $\text{FD}_{\text{PANNs}}$ $\downarrow$ & $\text{FD}_{\text{PaSST}}$ $\downarrow$ & $\text{KL}_{\text{PANNs}}$ $\downarrow$ & $\text{KL}_{\text{PaSST}}$ $\downarrow$ & $\text{IS}_{\text{PANNs}}$ $\uparrow$ & $\text{IS}_{\text{PaSST}}$ $\uparrow$ & $\text{CLAP}_\text{base}$ $\uparrow$ & $\text{CLAP}_\text{large}$ $\uparrow$ & $\text{LSD}$ $\downarrow$\\
    \midrule
    \multirow{3}{*}{\textbf{TTA}} & Joint Training & 15.865 & 144.478 & 1.314 & \textbf{1.133} & 11.923 & 9.515 & \textbf{0.513} & 0.413 & $-$ \\
    & Standard RoPE & 14.653 & 143.197 & \textbf{1.263} & 1.158 & 11.533 & 9.288 & 0.510 & 0.415 & $-$ \\
    & Ours & \textbf{14.288} & \textbf{134.341} & 1.282 & 1.171 & \textbf{12.168} & \textbf{9.734} & \textbf{0.513} & \textbf{0.418} & $-$ \\
    \midrule
    \multirow{3}{*}{\textbf{Editing}} & Joint Training & \textbf{10.942} & \textbf{106.642} & 0.510 & 0.379 & $-$ & $-$ & $-$ & $-$ & 1.326 \\
    & Standard RoPE & 11.168 & 107.931 & 0.505 & 0.384 & $-$ & $-$ & $-$ & $-$ & 1.324 \\
    & Ours & 10.991 & 106.664 & \textbf{0.484} & \textbf{0.361} & $-$ & $-$ & $-$ & $-$ & \textbf{1.321} \\
    \bottomrule
  \end{tabular}}
\end{table*}

\begin{table*}[!t]
    \caption{Ablation study on the guidance scale of CFG in TTA task.}
  \label{tab:TTAab}
    \centering
  \begin{tabular}{c|cc|cc|cc|cc}
    \toprule
    \textbf{CFG Setting} & $\text{FD}_{\text{PANNs}}$ $\downarrow$ & $\text{FD}_{\text{PaSST}}$ $\downarrow$ & $\text{KL}_{\text{PANNs}}$ $\downarrow$ & $\text{KL}_{\text{PaSST}}$ $\downarrow$ & $\text{IS}_{\text{PANNs}}$ $\uparrow$ & $\text{IS}_{\text{PaSST}}$ $\uparrow$ & $\text{CLAP}_\text{base}$ $\uparrow$ & $\text{CLAP}_\text{large}$ $\uparrow$\\
    \midrule
    {$\omega_0$ = 3.5} & 13.414 & 133.403 & 1.314 & 1.204 & 11.577 & 9.262 & 0.505 & 0.412 \\
    {$\omega_0$ = 4.0} & 13.807 & 133.858 & 1.295 & 1.186 & 11.916 & 9.525 & 0.509 & 0.415 \\
    {$\omega_0$ = 4.5} & 14.288 & 134.341 & 1.282 & 1.171 & 12.168 & 9.734 & 0.513 & 0.418 \\
    {$\omega_0$ = 5.0} & 14.631 & 135.303 & 1.277 & 1.162 & 12.442 & 9.926 & 0.516 & 0.421 \\
    {$\omega_0$ = 5.5} & 15.005 & 137.144 & 1.270 & 1.164 & 12.651 & 10.051 & 0.517 & 0.422 \\
    \bottomrule
  \end{tabular}
\end{table*}

\begin{table}[!t]
    \caption{Ablation study on the guidance scale of CFG in editing tasks}
  \label{tab:editab}
    \centering
  \resizebox{\columnwidth}{!}{
  \begin{tabular}{c|c|cc|cc|c}
    \toprule
    \textbf{$\boldsymbol{\omega_1}$} & \textbf{$\boldsymbol{\omega_2}$} &  $\text{FD}_{\text{PANNs}}$ $\downarrow$ & $\text{FD}_{\text{PaSST}}$ $\downarrow$ & $\text{KL}_{\text{PANNs}}$ $\downarrow$ & $\text{KL}_{\text{PaSST}}$ $\downarrow$ & $\text{LSD}$ $\downarrow$\\
    \midrule
    \multirow{3}{*}{1.0}
    & 3.0 & 10.891 & 104.645 & 0.479 & 0.358 & 1.331 \\
    & 4.0 & 11.011 & 105.031 & 0.481 & 0.356 & 1.332 \\
    & 5.0 & 11.048 & 106.290 & 0.482 & 0.356 & 1.335 \\
    \midrule
    \multirow{3}{*}{1.5}
    & 3.0 & 10.980 & 106.800 & 0.489 & 0.357 & 1.320 \\
    & 4.0 & 10.991 & 106.664 & 0.484 & 0.361 & 1.321 \\
    & 5.0 & 11.097 & 107.574 & 0.494 & 0.361 & 1.321 \\
    \midrule
    \multirow{3}{*}{2.0}
    & 3.0 & 11.064 & 108.580 & 0.509 & 0.368 & 1.314 \\
    & 4.0 & 11.066 & 109.241 & 0.510 & 0.369 & 1.316 \\
    & 5.0 & 11.126 & 110.149 & 0.505 & 0.369 & 1.319 \\
    \bottomrule
  \end{tabular}}
\end{table}

\subsection{Main Results}
\subsubsection{Text-to-Audio Generation}
In this section, we evaluate the TTA capability of our model and further investigate the impact of multistage training on TTA performance.

Table~\ref{tab:TTA} presents objective and subjective evaluation results on the TTA task. \textit{Ours-TTA-only} refers to the proposed model trained only on the TTA task, while \textit{Ours-Full} refers to the model trained through the full multistage training. \textit{Ours-TTA-only} demonstrate competitive performance across all objective metrics, achieving the best $\text{CLAP}_\text{large}$ scores. In subjective evaluation, it outperforms all baselines and achieves the second-best performance on both the MOS-Q and MOS-R metrics. \textit{Ours-Full} also shows strong performance on objective metrics, obtaining the second-best $\text{FD}_{\text{PANNs}}$ and $\text{CLAP}_\text{large}$ scores. For subjective metrics, it achieves the best human evaluation result on both MOS-Q and MOS-R.

Compared with \textit{Ours-TTA-only}, \textit{Ours-Full} achieves better FD, while its performance on other objective metrics declines. In contrast, both subjective evaluation metrics show improvements. We attribute this phenomenon to a shift in the data distribution induced by the introduction of audio editing data. Although this leads to a decrease in most objective metrics, from the perspective of human experience, the quality of the generated audio remains unaffected and even shows improved instruction following.

In summary, these experimental results demonstrate that our model achieves competitive performance on the TTA task. In addition, our multistage training strategy enables the model to acquire audio editing capabilities while preserving or even improving the TTA performance.

\subsubsection{Audio Editing}
In this section, we evaluate the performance of our model on six audio editing tasks and compare two approaches for reference audio conditioning. Furthermore, we study the effect of the fundamental generative capability of the model on its performance in editing tasks.

Table~\ref{tab:edit} summarizes objective and subjective evaluation results on six editing tasks. \textit{Ours-Edit-only} refers to the model employs our proposed architecture and is trained exclusively on editing tasks. DiT-Edit, \textit{Ours-Edit-only} and \textit{Ours-Full} outperform both pretrained baselines across all objective and subjective metrics. Compared with DiT-Edit which adopts a commonly used input-channel expansion design, \textit{Ours-Edit-only} achieves better overall performance on adding, reordering, inpainting and SR tasks, while remaining competitive results on removing and replacement tasks. These results demonstrate the effectiveness of our joint modeling approach in reference audio conditioning and instruction following.

\textit{Ours-Full} attains the best performance across all metrics except LSD on the inpainting task, indicating that the audio generation capability learned by the base model can benefit other generative tasks. In other tasks, \textit{Ours-Full} also achieves competitive results, demonstrating that its audio editing capability is comparable to task-specific models.

\subsection{Ablation Studies}
\subsubsection{Effectiveness of training strategy}
Table~\ref{tab:ablation} presents a comparison between our progressive multistage training strategy and a joint training strategy on both TTA and audio editing tasks. On the TTA task, our method achieves superior performance in all metrics except $\text{KL}_{\text{PaSST}}$. On audio editing tasks, it also outperforms joint training in all metrics except FD. These results suggest that our multistage training strategy yields more balanced performance on both TTA and editing tasks than joint training, especially mitigating the negative impact of task competition on TTA performance.

\subsubsection{Effectiveness of position embedding}
The comparison between the proposed factorized position embedding and the standard Rotary Position Embedding (RoPE) is also presented in Table~\ref{tab:ablation}. On the TTA task, our position embedding achieves superior performance across all metrics except KL divergence. For editing tasks, our position embedding consistently outperforms standard RoPE across all metrics. These results suggest that our factorized position embedding is more capable of capturing both inter-sequence dependencies and intra-sequence positional relationships simultaneously.

\subsubsection{CFG Scale}
We conduct comparative experiments with different classifier-free guidance (CFG) scales. As shown in Table~\ref{tab:TTAab}, we evaluate the impact of the guidance scale $\omega_0$ in~(\ref{eq:ttacfg}) on the performance of the TTA task. Within the range of the guidance scale from 3.5 to 5.5, FD generally increases as the guidance scale increases, while the KL divergence shows a decreasing trend, and both IS and CLAP scores tend to increase with a higher guidance scale. These results indicate a trade-off between distribution fidelity and both semantic and quality performance, where increasing the guidance scale improves KL, IS, and CLAP scores at the expense of higher FD.

As presented in Table~\ref{tab:editab}, we evaluate the impact of the guidance scale $\omega_1$ and $\omega_2$ in~(\ref{eq:mtcfg}) on the performance of the model in editing tasks. In the same setting of $\omega_2$, the LSD decreases as $\omega_1$ increases, indicating that the generated audio better matches the target in the frequency domain. When $\omega_1$ is set to 1.0 or 2.0, increasing $\omega_2$ from 3.0 to 5.0 leads to degradation in both FD and LSD. When $\omega_2$ is fixed to 5.0, both FD and KL deteriorate as $\omega_1$ increases. Setting $\omega_1$ to 1.5 and $\omega_2$ to 3.0 or 4.0 achieves competitive performance across all metrics. These results suggest that moderate and balanced guidance scales can provide an appropriate trade-off between spectral consistency and perceptual quality in editing tasks.

\section{Conclusion}
\label{sec:conclusion}
In this paper, we propose AudioWeave, a unified model for audio generation and editing that operates with shared model architecture and parameters without requiring additional task-specific components. We employ a hybrid diffusion transformer backbone composed of dual-stream MMDiT blocks and single-stream DiT blocks to achieve a balance between model performance and parameter efficiency. Through the proposed joint condition modeling with factorized position embedding, our model enables comprehensive interactions among text, target audio, and optional reference audio. Due to the varying input conditions and data distributions across different tasks, we introduce a progressive multistage training strategy. Experimental results demonstrate that our model achieves performance comparable to that of task-specific models on both TTA and audio editing tasks. In future work, we will extend our model to incorporate a broader range of generative tasks and enable it to handle multiple reference audio inputs simultaneously.

\bibliographystyle{IEEEtran}
\bibliography{reference}

\vfill

\end{document}